\documentclass[10pt, aps, prx, twocolumn,show pacs,show keys,superscript address,superscript reference,floatfix, ]{revtex4-2}
\usepackage[colorlinks=true,citecolor=blue,linkcolor=blue,urlcolor=blue]{hyperref}
\usepackage{graphicx}
\usepackage{color}
\usepackage{amssymb}

\usepackage{subeqnarray}
\usepackage{float}
\usepackage{bbm}
\usepackage{bm}
\usepackage{diagbox}
\usepackage{makecell}
\usepackage[T1]{fontenc}
\usepackage{selinput}
\usepackage{babel}
\usepackage{amsmath}
\usepackage{tabularx, booktabs,lipsum}
\usepackage{soul}
\usepackage{lineno}

\DeclareMathAlphabet\mathbfcal{OMS}{cmsy}{b}{n}
\newcommand{\ket}[1]{\ensuremath{\left|{#1}\right\rangle}}

\begin{document}
\title{Digital-analog quantum computing of fermion-boson models in superconducting circuits}
\author{Shubham~Kumar\thanks{Corresponding author: shubhamkumar.kumar@gmail.com}}
\email{shubhamkumar.kumar@gmail.com}
\affiliation{Kipu Quantum, Greifswalderstrasse 212, 10405 Berlin, Germany}

\author{Narendra~N.~Hegade }
\affiliation{Kipu Quantum, Greifswalderstrasse 212, 10405 Berlin, Germany}

\author{Anne-Maria Visuri}
\affiliation{Kipu Quantum, Greifswalderstrasse 212, 10405 Berlin, Germany}

\author{B.~A.~Bhargava}
\affiliation{Kipu Quantum, Greifswalderstrasse 212, 10405 Berlin, Germany}

\author{Juan~F.~R.~Hernandez}
\affiliation{Kipu Quantum, Greifswalderstrasse 212, 10405 Berlin, Germany}

\author{E.~Solano}
\affiliation{Kipu Quantum, Greifswalderstrasse 212, 10405 Berlin, Germany}

\author{F.~Albarr\'an-Arriagada}
\affiliation{Departamento de F\'isica, CEDENNA, Universidad de Santiago de Chile (USACH), Avenida V\'ictor Jara 3493, 9170124, Santiago, Chile.}

\author{G.~Alvarado~Barrios}
\affiliation{Kipu Quantum, Greifswalderstrasse 212, 10405 Berlin, Germany}

\begin{abstract}

High-fidelity quantum simulations demand hardware-software co-design architectures, which are crucial for adapting to complex problems such as strongly correlated dynamics in condensed matter. By leveraging co-design strategies, we can enhance the performance of state-of-the-art quantum devices in the noisy intermediate quantum (NISQ) and early error-correction regimes. In this direction, we propose a digital-analog quantum algorithm for simulating the Hubbard-Holstein model, describing strongly-correlated fermion-boson interactions, in a suitable architecture with superconducting circuits. It comprises a linear chain of qubits connected by resonators, emulating electron-electron (e-e) and electron-phonon (e-p) interactions, as well as fermion tunneling. Our approach is adequate for digital-analog quantum computing (DAQC) of fermion-boson models, including those described by the Hubbard-Holstein model. We show the reduction in the circuit depth of the DAQC algorithm, a sequence of digital steps and analog blocks, outperforming the purely digital approach. We exemplify the quantum simulation of a half-filled two-site Hubbard-Holstein model. In this example, we obtain time-dependent state fidelities larger than 0.98, showing that our proposal is suitable for studying the dynamical behavior of solid-state systems. Our proposal opens the door to computing complex systems for chemistry, materials, and high-energy physics. 

\end{abstract}

\maketitle

In solid-state physics, the Hubbard-Holstein (HH) model is commonly used to describe fermionic lattices interacting with phonons~\cite{Berger1995PhysRevB}. This model captures a range of phenomena, such as the formation of polarons---quasiparticles consisting of an electron or hole dressed by a phonon cloud---and their bound states~\cite{Fehske.1996, Ventriglia.2001,Martin.2016}. It plays a vital role in the determination of the transport properties of biomolecules~\cite{Lionel.1965, Yuri.2007}, as well as the correlation effects and localization phenomena in materials such as cuprates, fullerides, and manganites~\cite{Novelli.2014,Nagaosa.2004, Kabanov.1998, Esposito.2017}. Consequently, the HH model and similar systems, describing fermionic lattices coupled to bosons, are fundamental in various research fields and technologies in chemistry, materials, and high-energy physics.

Strongly correlated fermionic systems, including those involving fermion-phonon interactions, pose significant challenges due to their complexity. Achieving exact solutions for general fermionic systems becomes impractical beyond a few particles, leading to the need for approximations and symmetry conditions. Various theoretical and computational techniques have been developed to address these challenges, including mean-field theories~\cite{JimenezHoyos2015PhysRevB,Jeon2004PhysRevB,Backes2023PhysRevB}, quantum Monte Carlo methods~\cite{Johnston2013PhysRevB,Costa2020CommunPhys}, density functional theory~\cite{Malet2015PhysRevLett}, tensor network methods~\cite{Corboz2010PhysRevB, Cirac2021RevModPhys}, and the density matrix renormalization group algorithm~\cite{Jeckelmann1998PhysRevB, Schollwoeck2011AnnalsofPhysics}. These approaches provide ways to approximate the behavior of many-body systems and gain insights into their properties, although mean-field theories overlook important correlation effects. These techniques play a crucial role in advancing technology in the era of the second quantum revolution.

Quantum simulation and quantum computing offer advanced approaches for studying the dynamics of strongly-correlated systems~\cite{Nori.2014}. By encoding the targeted model into controllable quantum processors, the quantum evolution can be mimicked in a digital, analog, or digital-analog manner. In digital simulations, a specific sequence of single-qubit and two-qubit gates, forming a universal set of quantum logic gates, is used to emulate any quantum evolution. However, the fidelity of single and two-qubit gates, as well as qubit coherences, limit the usefulness of this approach in our noisy intermediate-scale quantum (NISQ) era. In contrast, analog simulations naturally perform the desired quantum evolution within specific parameter ranges with a high accuracy, but it only works for the proposed specific model. Examples include quantum simulations of light-matter interactions in trapped ions~\cite{Lv2018PhysRevX,Pedernales2015SciRep} and superconducting circuits \cite{FengPRB2013}. A merged approach called digital-analog quantum computing (DAQC)~\cite{ Mikel.2020} bridges the gap between these two paradigms. DAQC leverages analog dynamics from the simulator Hamiltonian, in the spirit of analog blocks, which are combined with digital steps, using the Trotter-Suzuki decomposition in an algorithmic manner. In this \textit{co-design} approach, the quantum algorithms and hardware architecture are developed together, tailoring each to optimize the overall system performance \cite{Anna.2023}.

\begin{figure*}
    \centering
    \includegraphics[width=0.95\linewidth]{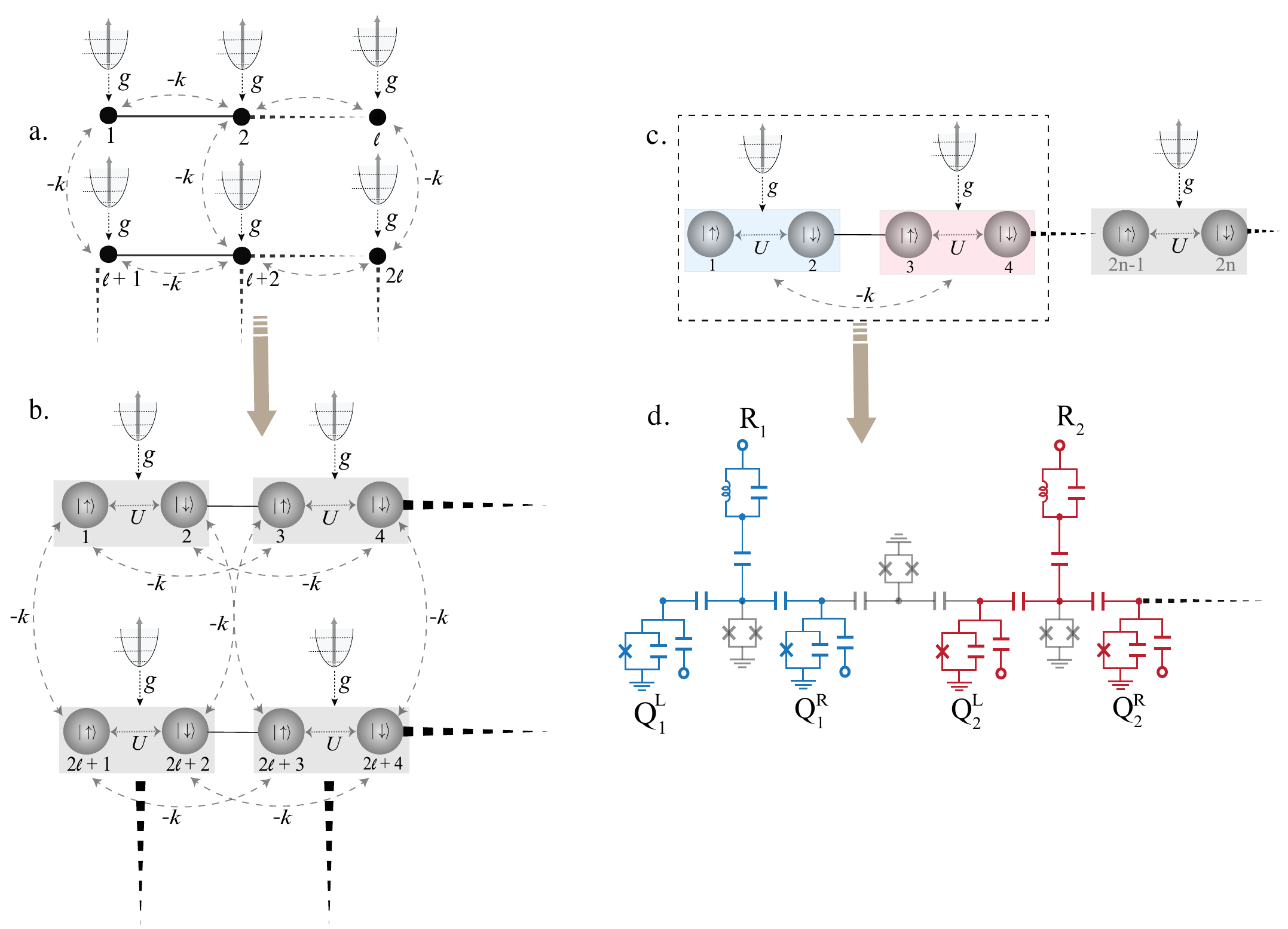}
    \caption{a. The HH lattice to be simulated. The bosonic frequency is $\omega_0$, the hopping energy between lattice sites is $-k$, and $g$ is the e-p coupling strength. b. Mapping of HH lattice to an equivalent spinless fermion lattice using the JW transformation. Each site (shaded) is represented by two fermionic pairs and the on-site repulsion is given by $U$.  c. Representation of the mapped 1-D HH model (each site with different color). d. The corresponding superconducting circuit architecture representing the spin-boson chain using a 1-D linear chain of transmon qubits (Q$_j^{L(R)}$) coupled via resonators (R$_j$). }
    \label{Fig01}
\end{figure*}

Quantum simulations have been proposed for studying strongly-correlated fermionic systems, including e-p interactions, on various platforms such as trapped ions~\cite{Solano.2012}, cold atoms~\cite{Hague.2012,Leticia.2018,Hofstetter2018}, and superconducting circuits~\cite{Solano.2016, Lamata.2015}. For qubit-based platforms, the key challenge lies in mapping fermionic degrees of freedom onto spin degrees of freedom, achieved for example through the Jordan-Wigner transformation. Trapped ions benefit from the M{\o}lmer-S{\o}rensen gate, whether as a well-behaved two-qubit gate or a multiqubit quantum operation. Alternatively, superconducting architectures have also been advanced, demonstrating comparable scalability features through parallel interactions and geometric considerations~\cite{Jing.2021}. Moreover, superconducting circuits offer an ideal platform for compact digital-analog quantum simulations of strongly-correlated fermionic systems with multiple bosonic modes, like the Hubbard-Holstein model. Such versatility is due to the possible addition of multiple waveguide cavities or LC oscillators.

We present a superconducting circuit design for compact simulation of the HH model within the DAQC paradigm, applicable to one- and two-dimensional arrays. Our design comprises qubits, resonators, and superconducting quantum interference devices (SQUIDs) arranged along a chain. This architecture enables natural qubit-resonator and tunable qubit-qubit interactions, facilitating the implementation of analog blocks to encode the evolution under the HH model. Such hybrid continuous-variable and discrete-variable quantum computing hardware combines the strengths of both systems, enabling advanced quantum computations and error correction, and is supported by novel compilation techniques and formal architectures to bridge applications and hardware \cite{Girvin.2024}. We employ the DAQC protocol on a toy model of a half-filling two-site HH Hamiltonian, numerically calculating the fidelity respect to the exact evolution. Comparing our results to a purely digital approach, we obtain an enhanced accuracy. Furthermore, by evaluating resource scaling in terms of gate count, we surpass the performance of existing trapped-ion platforms. Our proposal is thus relevant for exploring phase diagrams, polaron formation, and applications in materials, with implications in chemistry and high-energy physics.

\section{Results} The HH model is a minimal description of how electrons on a lattice interact due to their electric charge and couple to phononic vibrations. The model includes e-p interactions, e-e repulsion, and electron hopping between lattice sites. For an $N$-site lattice, it is described by the Hamiltonian 
\begin{eqnarray}
H&=&\omega_0 \sum_{j=1}^Na_{j}^{\dagger} a_{j}+U\sum_{j=1}^N(\hat{n}_{j \uparrow} \hat{n}_{j \downarrow}) - k \sum_{\langle j,l \rangle}\sum_{\sigma=\uparrow,\downarrow} c_{j, \sigma}^{\dagger} c_{l, \sigma}^{\phantom{\dagger}} \nonumber \\ 
&&+  g\sum_{j=1}^N\sum_{\sigma=\uparrow,\downarrow}[\hat{n}_{j,\sigma}(a_{j}^{\dagger}+a_{j})],
\label{Eq01}
\end{eqnarray}
where $c_{j \sigma}^{\dagger} (c_{j \sigma}^{\phantom{}})$ is the fermionic creation (annihilation) operator for site $j$ and spin $\sigma$, $a_j^{\dagger} (a_j^{\phantom{}})$ is the creation (annihilation) operator for the $j$th bosonic mode, and $\hat{n}_{j \sigma} = c_{j \sigma}^{\dagger} c_{j \sigma}^{\phantom{}}$. The parameters $\omega_0$ and $U$ are the bosonic frequency and on-site interaction, respectively. From the coupling terms, $k$ is the hopping energy in the lattice, $\langle j,l\rangle$ refers to nearest-neighbour sites, and $g$ represents the e-p coupling strength. We note that if $\omega_0=g=0$ (without bosonic modes), the resulting Hamiltonian corresponds to the Hubbard model. On the other hand, if $U=0$ (fermions without charge), we obtain the Holstein model.

To simulate the HH model, we need to encode the bosonic and fermionic degrees of freedom. For the first one, in superconducting circuits, we can use transmission lines or $LC$ oscillators, obtaining an analog to the phonons in the HH model. On the other hand, the fermion operators (with spin) can be mapped to spinless fermion operators in a lattice of double the size, that is $c_{j,\uparrow}\rightarrow b_{2j}$ and $c_{j,\downarrow}\rightarrow b_{2j+1}$. Finally, to encode spinless fermionic operators, we perform the Jordan-Wigner (JW) transformation \cite{Jordan.1928, Aspuru.2020}, $b_{j} \rightarrow \prod_{j=1}^{i-1} (-\sigma_{j}^{z}) \sigma_{i}^{-}$. In this manner, we map the HH lattice to a spin chain interacting with bosonic modes (see Fig.~\ref{Fig01}). The latter can be described by the Hamiltonian

\begin{align}
H = & \sum_{j=1}^N \Big[\omega_0 a_{j}^{\dagger} a_{j}+\frac{\bar{U}}{4}(\sigma_{2j}^z+\sigma_{2j-1}^z)+\frac{U}{4}(\sigma_{2j}^z\sigma_{2j-1}^z)\nonumber \\ 
&+  \frac{g}{2}(\sigma_{2j}^z+\sigma_{2j-1}^z)(a_{j}^{\dagger}+a_{j})\Big]\nonumber\\
&- k \sum_{\langle j,l\rangle}\Bigg[\sigma_{2l}^-\bigg(\prod_{m=2l}^{2j-1}-\sigma_m^z\bigg)\sigma_{2j}^+ \nonumber \\ 
&  + \sigma_{2l-1}^-\bigg(\prod_{m=2l-1}^{2j-2}-\sigma_m^z\bigg)\sigma_{2j-1}^+ + {\rm H.c.} \Bigg] ,
\label{Eq04}
\end{align}
where the term $g(a_j+a_j^{\dagger})$ was absorbed by the change of variables $a_j\rightarrow a_j+g/\omega_0$ and therefore $\bar{U}=U-4g^2/\omega_0$.
We can see that the first and second term in Eq.~(\ref{Eq04}) correspond to the free energy of bosonic and spin degrees of freedom, easily implementable by a set of $LC$ oscillators (or transmission lines) and qubits. The next two terms represent the qubit-qubit and qubit-boson interactions, which can be done in superconducting circuits using capacitive coupling and local rotations in the qubits. The last terms correspond to the fermionic tunneling terms, which can be implemented by using consecutive two-body rotations, as shown in~\cite{Jing.2021}.

\subsection{Digital analog quantum simulator} 
We describe here an architecture that allows to implement the Hamiltonian of Eq.~(\ref{Eq04}) efficiently: a chain of qubits coupled to a set of resonators. To simulate this Hamiltonian, we propose the superconducting circuit architecture shown in Fig.~\ref{Fig01}d. Such a superconducting design consists of a linear array of transmon qubits (Q$_j^{L(R)}$, where $j$ is the site index and L(R) denotes the left or right qubit at the same site) coupled through grounded superconducting quantum interference devices (SQUIDs). In addition, the chain of qubits has LC resonators (R$_j$) coupled at each site by grounded SQUIDs. For the transmon array, we intersperse devices with different spectra. It means that nearest-neighbour qubits are off resonance, \textcolor{blue}{i.e.} qubits in odd positions have a frequency different from qubits in even positions, and the resonators also have a different frequency \textcolor{blue}{than} the qubits. Therefore, our architecture can be seen as a chain of building blocks, enclosed by a dashed square in Fig.~\ref{Fig01}, composed of three different devices: a left qubit with frequency $\omega_L$, a right qubit with frequency $\omega_R$, and a resonator with frequency $\omega$. A similar architecture without resonators was proposed in~\cite{Jing.2021}, and another one including experimentally performed resonators in~\cite{Schoelkopf.2020}, both in the context of quantum simulations.

The interactions between qubits and between each qubit and the resonator at each site are controlled through a single SQUID. This is achieved by applying an external flux to the SQUID, consisting of a DC component and a small AC signal. The flux is expressed as (see Supplementary Note A)
\begin{equation*}
\varphi_{ext}=\varphi_{DC}+\varphi_{AC}(t),
\end{equation*}
where 
\begin{equation*}
\varphi_{AC}(t) = \varphi_{R} + \sum_{\alpha}\varphi_{QQ}^{\alpha} + \varphi_{QR}^{\alpha(1)} + \varphi_{QR}^{\alpha(2)}.
\end{equation*}
Here, $\alpha=\{+,-\}$ and $\varphi_{ab}^{\alpha} = A_{ab}\cos(\nu_{ab}^{\alpha}t + \tilde{\varphi}_{ab}^{\alpha})$ with the amplitude $A_{ab}$, frequency $\nu_{ab}$ and phases $\tilde{\varphi}_{ab}^{\alpha}$. With this form of the signal, we specify which part of the subsystem is controlled by $\tilde{\varphi}_{ab}^{\alpha}$. We can then turn on (off) the specific phases $\tilde{\varphi}_{ab}^{\alpha}$ to activate (deactivate) the couplings. The full evolution of the qubit-qubit and qubit-resonator interactions is achieved in a sequence of Trotter steps.

After the high-plasma-frequency and low-impedance-SQUID approximations, our architecture is described by the interaction-picture Hamiltonian
\begin{equation}
H=\sum_{j=1}^{N}H_{s}^{(j)} + \sum_{j=1}^{N-1}H_{s-s}^{(j)} \, ,
\label{Eq05}
\end{equation}
where $H_{s}^{(j)}$ and $H_{s-s}^{(j)}$ are the $j$th building-block Hamiltonian and the interaction between building blocks $j$ and $j+1$, respectively. Such Hamiltonians have the form (for a detailed calculation see the Supplementary Note A)
\begin{widetext}
\begin{align}
H_{s}^{(j)} = & \frac{m_1 A_{QQ}}{4} \Big[ C_{QQ}^+ \sigma_{L,j}^x \sigma_{R,j}^x - S_{QQ}^+ \sigma_{L,j}^x \sigma_{R,j}^y + S_{QQ}^- \sigma_{L,j}^y \sigma_{R,j}^x + C_{QQ}^- \sigma_{L,j}^y \sigma_{R,j}^y \Big] \nonumber \\
& + \frac{1}{4} \sum_{\alpha=\{L,R\}} A_{QR} l_{1}^{\alpha} \Big[ C_{QR}^+ \sigma_{\alpha,j}^x (a_j^{\dagger} - a_j) - C_{QR}^- \sigma_{\alpha,j}^y (a_j^{\dagger} + a_j) 
+ S_{QR}^+ \sigma_{\alpha,j}^x (a_j^{\dagger} + a_j) - S_{QR}^- \sigma_{\alpha,j}^y (a_j^{\dagger} - a_j) \Big], \nonumber \\
H_{s-s}^{(j)} &= \frac{m_1 \bar{A}_{QQ}}{4} \Big[ C_{\phi QQ}^+ \sigma_{R,j}^x \sigma_{L,j+1}^x - S_{\phi QQ}^+ \sigma_{R,j}^x \sigma_{L,j+1}^y+ S_{\phi QQ}^- \sigma_{R,j}^y \sigma_{L,j+1}^x + C_{\phi QQ}^- \sigma_{R,j}^y \sigma_{L,j+1}^y \Big]
\label{Eq06},
\end{align}
\end{widetext}
where $C_{ab}^{\pm} = \cos \tilde{\varphi}_{ab}^+ \pm \cos \tilde{\varphi}_{ab}^-$, $\tilde{C}_{ab}^\pm = \cos\tilde{\phi}_{ab}^+ \pm \cos\tilde{\phi}_{ab}^-$, and similar definitions for $S_{ab}^{\pm}$ and $S_{ab}^{\pm}$, replacing $\cos$ by $\sin$. The phases $\tilde{\varphi}_{ab}^{\pm}$ are associated with the magnetic flux through the SQUIDs inside each building block, and $\tilde{\phi}_{ab}^{\pm}$ are the phases of the magnetic flux through the SQUIDs that couple different building blocks. The magnetic flux parameters required to activate or deactivate specific interactions are shown in Supplementary Table 1. We note that by adjusting the different phases, we can engineer a wide range of qubit-qubit and qubit-boson interactions.

\begin{figure*}
\centering
\includegraphics[width=1\linewidth]{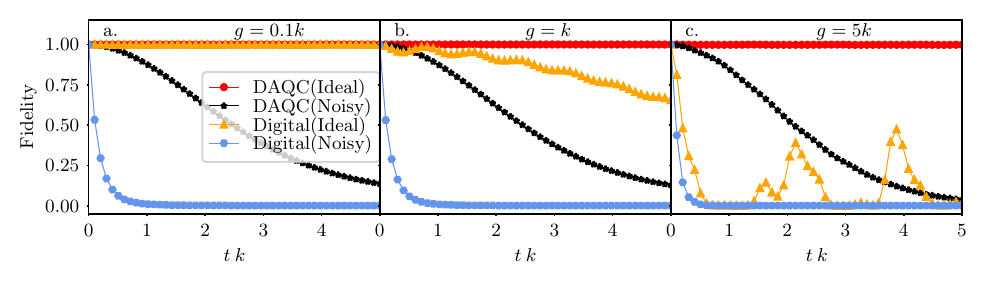}
\caption{Time-dependent state fidelity $|\langle\psi_{exact} | \psi_{sim}\rangle|^2$ with and without noise as a function of time for various parameters. The initial state is half-filling, $\psi_i = \ket{0,1,1,0,0,0}$. Each resonator was truncated to eight levels.}
\label{fig.3}
\end{figure*}


\begin{figure}
\centering
\includegraphics[width=0.95\linewidth]{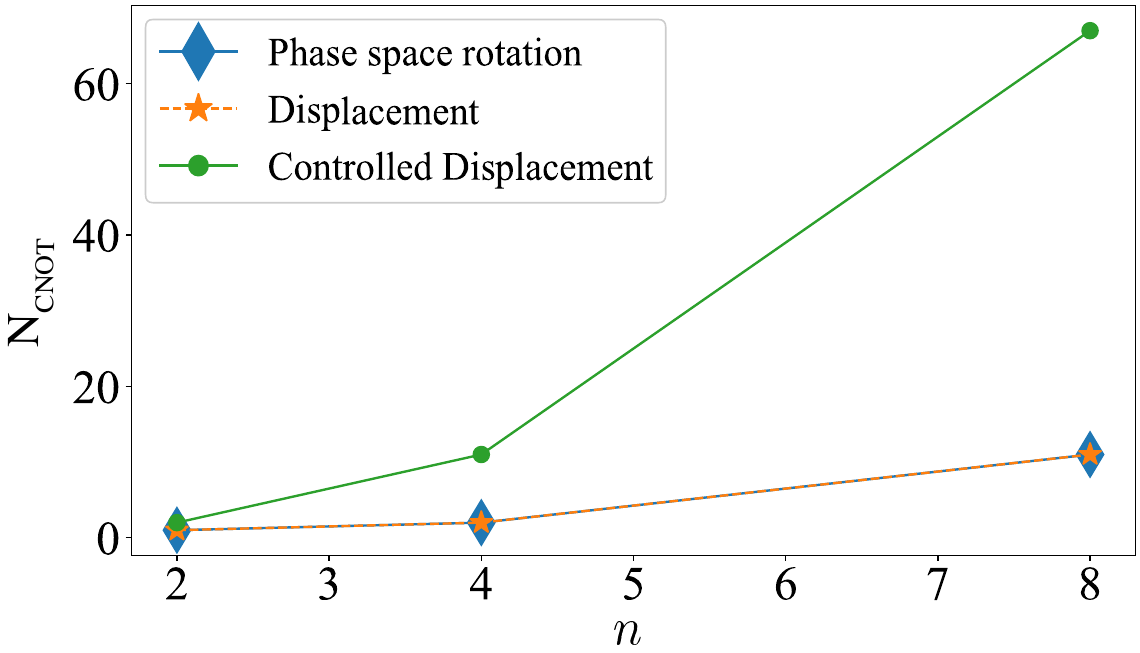}
\caption{Number of CNOTs ($\rm{N_{CNOT}}$) required to perform each bosonic gate as a function of the number of levels ($n$) for each bosonic mode for a pure digital approach.}
\label{fig.5}
\end{figure}
\begin{figure*}
\centering
\includegraphics[width=1.0\linewidth]{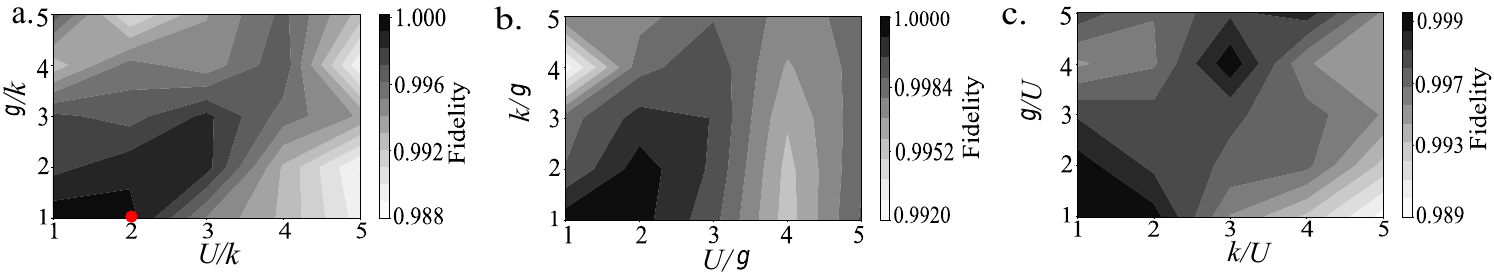}
\caption{DAQC fidelity $\vert \langle\psi_{exact} | \psi_{sim}\rangle \vert^2$ as a function of a. the on-site interaction $U$ and the e-p coupling $g$ in units of the hopping energy $k$. The red dot represents the last point of the ideal DAQC simulation curve in Fig. 3 (red). b. $U$ and $k$ in units of $g$, and c. $k$ and $g$ in units of $U$ for 50 Trotter steps. The initial state is half-filling, $\psi_i = \ket{0,1,1,0,0,0}$, where each resonator was truncated to 8 levels.}
\label{fig.6}
\end{figure*}

\subsection{Simulations} For simulation, we consider the dimensionless Hamiltonian $H/k$ and choose the timescale in units of $1/k$. For different materials, the range of parameters of the HH model of Eq.~(\ref{Eq01}) is such that $k$ and $U$ lie between $0.2-10$ eV, $g$ is between $0-100$ eV and $\omega$ lies in the range $0.01-1$ eV \cite{Pai.2015}. The corresponding physical time scales of the system are between fs and ps. Considering a microwave driving with qubit and resonator frequencies in the range $5-10$ GHz, the Trotter time step for the digital blocks comprising the single and two qubit gates are in the range $10-500$ ns \cite{Wendin.2017}, while for the analog blocks comprising the bosonic and qubit-bosonic gates, they are around $72$ ns \cite{Schoelkopf.2020}. The initial states of the system are defined using the ordering of the Hilbert space as $\ket{Q_1, Q_2, Q_3, Q_4, res_1, res_2}$, where $Q_i$ and $res_i$ denote the qubit and resonator subspaces, respectively. 


In Fig.~\ref{fig.3}, we show the time-dependent state fidelity $|\langle\psi_{exact}|\psi_{sim}\rangle|^2$ of the noise-based simulation as a function of time for 50 Trotter steps. The time step $\delta t$ is kept constant, so that the total evolution consists of 50 discrete steps. The noisy DAQC simulation incorporates amplitude-damping errors for both the resonators and the qubits. The damping factors are determined based on the duration of the analog and digital blocks $T_{\text{block}}$ divided by their relaxation times $T_1$ of the resonators and qubits, such that $\gamma = T_{\text{block}}/T_1$. The duration of the analog block is $50 \, \text{ns}$, the duration of the digital block is $200 \, \text{ns}$, and the relaxation time for both qubits and resonators was set to $T_1 = 80 \mu s$. This leads to calculated damping factors of $\gamma_{\text{res}} = 0.000625$ for the resonators and $\gamma_{\text{qubit}} = 0.0025$ for the qubits. 

Unlike digital blocks, which accumulate errors from both amplitude damping and gate imperfections, analog blocks are only affected by amplitude damping without additional noise sources introduced by discrete gate operations. Noise is applied using Kraus operators to model amplitude damping, where each Kraus operator $A_k(t)$ is constructed for a given time $t$, decay rate $\gamma$, and system dimension. These operators modify the quantum state iteratively by applying them in time steps, ensuring that the normalization condition $\sum_k A_k^\dagger(t) A_k(t) = I$ is satisfied to conserve total probability. For the reservoir components, noise is applied using a set of tensor product Kraus operators specific to the subsystem dimensions, while for the qubit components, similar noise channels target individual qubits. In the noisy digital simulation, depolarizing errors are combined with thermal relaxation errors for each gate, with gate fidelities derived from IBM devices. The digital noise model uses \textit{Qiskit} noise model to construct these errors, handling both single-qubit and two-qubit operations by combining depolarizing and thermal relaxation components to reflect realistic noise conditions in digital quantum simulations. 

The oscillations in the fidelity curves of the ideal cases, shown by the orange triangles (Digital) in Fig.~\ref{fig.3}, and the red dots (DAQC) in the same figure, appear from the non-commuting terms in the Hamiltonian, for which the Trotter decomposition of the unitary time evolution operator introduces a phase factor. Such oscillations have been documented in previous studies \cite{Solano.2012, Julen.2016}. In the noisy simulations, it is observed that noise affects these oscillations and they are canceled out. The reduced oscillation amplitude in the DAQC approach is expected as fewer Trotter decompositions are performed. The DAQC approach, therefore, achieves more stable dynamics with fewer artifacts, making it a powerful method for simulating complex quantum systems.

The digital simulations were carried out using \textit{Bosonic Qiskit} \cite{Wiebe.2022}, where each resonator was truncated to $n$ levels. This requires an order of $\log_2 n$ number of digital gates each, apart from the gates required for the qubits. The terms $a^{\dagger}a$, $a^{\dagger}+a$, and $\sigma_z(a^{\dagger} + a)$ can be implemented using cv\_r$()$, cv\_d$()$, and cv\_c\_d$()$ gates, respectively. This is provided by \textit{Bosonic Qiskit} and corresponds to the phase space rotation, displacement, and controlled displacement operators, respectively. When decomposed into digital gates, these bosonic gates can be realized using combinations of single-qubit rotations $U3 (\theta, \phi, \lambda)$ and CNOTs. 

Since entangling gates are common in NISQ processors, we estimate the number of CNOTs required for each of these bosonic gates for a single Trotter step. We plot the CNOTs required per bosonic gate with increasing resonator levels ($n$) in Fig.~\ref{fig.5}. The CNOTs required for the gates, cv\_r$()$ and cv\_d$()$, grow equally. In case of cv\_c\_d$()$, it grows with a very high margin requiring a total of 67 CNOTs for $n=8$.

From a dynamical point of view, our DAQC protocol can produce very high fidelities. For the two-site HH model, we compute the overlap between the exact solution and the numerical simulation using the DAQC protocol for different regimes as shown in Fig.~\ref{fig.6}. Based on the observation that a high-fidelity simulation of the HH model is feasible, as seen in Fig.~\ref{fig.3}, we investigate which coupling parameters (\(k\), \(g\), and \(U\)) can be realized by the superconducting architecture to accurately simulate the model in these regimes. The coupling regimes suitable for a two-site simulation are presented in Fig.~\ref{fig.6} (ideal simulation). In Fig.~\ref{fig.6} a., the red marker indicates the final fidelity of the ideal DAQC simulation shown in Fig.~\ref{fig.3}. The range of fidelities in Fig.~\ref{fig.3} is above 0.988,  which is consistent with the ideal DAQC curve in Fig.~\ref{fig.3}. The architecture simulates the two-site HH model efficiently within the given parameter regime and can therefore be extended to a linear-chain or a 2-D architecture. For the latter, there is a possibility to demonstrate quantum advantage for interacting fermion-boson systems. We observe that the fidelity \textcolor{blue}{in Fig.~\ref{fig.6}} is higher than 0.98, making our proposal suitable for studying dynamical properties of condensed matter systems. Moreover, since the scaling of our proposal depends only on the smaller dimension of the lattice, we anticipate that these results can be maintained for larger systems.

\subsection{Quantum simulation of condensed-matter physics}
The formation and decay of bound states due to electron-electron and electron-phonon interactions have profound effects on the electrical and thermal transport properties of materials~\cite{franchini2021polarons}. Despite its simplicity, the HH model captures some of the important non-trivial phenomena arising from these interactions.
The DAQC protocol proposed here can be used to simulate the real-time dynamics of this model, not easily accessible in condensed-matter experiments. For instance, spatially-resolved and temporally-resolved density-density correlation functions, accessible here, were used in previous numerical studies to characterize polaron formation~\cite{vidmar2011nonequilibrium, golez2012dissociation}. Exotic bound states such as repulsively bound pairs could also be simulated---these states can occur in lattice systems when the interaction energy is larger than the kinetic energy bandwidth. They are expected to decay quickly due to dissipation but were recently observed in antiferromagnetic compounds~\cite{wang2024experimental} and in cold atoms in optical lattices~\cite{winkler2006repulsively}. 

As an example of the physical quantities accessible with the proposed hardware and algorithm, we compute the electron double occupation $\langle \hat{n}_{j \uparrow} \hat{n}_{j \downarrow} \rangle$ and the total number of phonons $\langle \hat{n}_{ph} \rangle = \langle \hat{a}_2^{\dagger} \hat{a}_2^{\phantom{\dagger}} \rangle + \langle \hat{a}_2^{\dagger} \hat{a}_2^{\phantom{\dagger}} \rangle$ as functions of time. These quantities characterize the dynamics of a repulsively bound pair (with $U \gg k$) initially localized on site $j = 1$. We consider a weak and a strong electron-phonon coupling in the regime $g\leq 5k$.

\begin{figure}
\centering
\includegraphics[width=\columnwidth]{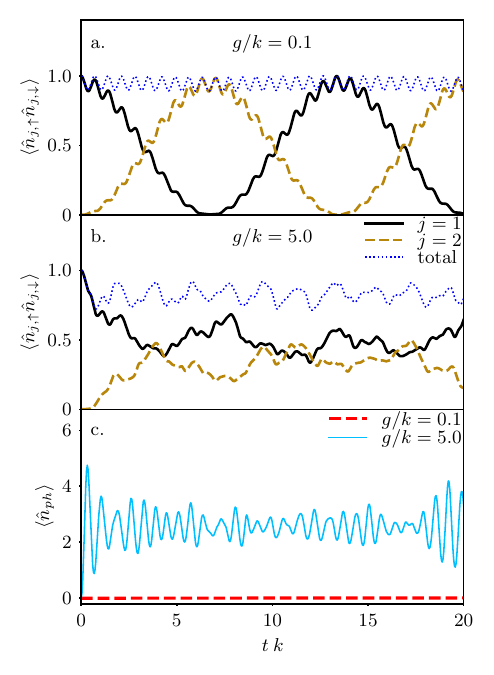}
\caption{a. The electron double occupation $\langle \hat{n}_{j \uparrow} \hat{n}_{j \downarrow} \rangle$ on sites $j = 1$ and $j = 2$ and the total double occupation $\langle \hat{n}_{1 \uparrow} \hat{n}_{1 \downarrow} \rangle + \langle \hat{n}_{2 \uparrow} \hat{n}_{2 \downarrow} \rangle$ computed with the DAQC algorithm. These quantities show coherent oscillations as functions of time for weak coupling $g = 0.1 k$. b. For strong coupling $g = 5k$, the oscillations become irregular and the average total double occupation is reduced. c. The total number of phonons $\langle \hat{n}_{ph} \rangle = \langle \hat{a}_1^{\dagger} \hat{a}_1^{\phantom{\dagger}} \rangle + \langle \hat{a}_2^{\dagger} \hat{a}_2^{\phantom{\dagger}} \rangle$ is close to zero for weak coupling. For strong coupling, it is nonzero with irregular oscillations due to the large number of degrees of freedom involved in the dynamics. Here, the on-site interaction and phonon frequency are $U = \omega_0 = 8k$.}
\label{Fig07}
\end{figure}

For weak coupling $g = 0.1 k$, shown in Fig.~\ref{Fig07}(a), the total electron double occupation $\langle n_{1 \uparrow} n_{1 \downarrow} \rangle + \langle n_{2 \uparrow} n_{2 \downarrow} \rangle$ stays close to the initial value~$1$, indicating that the electrons remain in a bound state. The high-frequency oscillation (with frequency close to $U$) results from a virtual pair-breaking and reassociation process where one electron tunnels to the second site and back.
The site-resolved double occupation also shows a slower oscillation associated with the tunneling of the bound pair between the two sites. The total number of phonons in Fig.~\ref{Fig07}(d) is close to zero in this case. For $g = 5k$ [Fig.~\ref{Fig07}(b)], the oscillations become irregular as the strong coupling to phonons introduces more degrees of freedom and more eigenfrequencies contribute to the dynamics. The total double occupation is slightly reduced and the number of phonons is nonzero, which suggests that the pair can be broken when the interaction energy is dissipated by the creation of phonons. These observations illustrate the possibilities of the DAQC algorithm for the quantum simulation of condensed-matter physics relevant for material properties. They also highlight its advantage over a fully digital algorithm, since a high fidelity is crucial for the measurement of high-frequency oscillations.



\section{Discussions} We developed a co-design DAQC approach for the quantum simulation of strongly-correlated fermion-boson models in superconducting circuits. The proposed digital-analog encoding allows us to simulate the problem Hamiltonian with high fidelity and fewer resources as compared to the purely-digital case. We obtain a circuit-depth scaling proportional to $\ell^2$ for a $\ell\times h$ lattice where $\ell\le h$. Such a compact scaling cannot be obtained with purely digital quantum computing, where the number of controlled gates and qubits grows with both dimensions in the lattice. Conversely, the scaling provided by other platforms like trapped ions grows with both dimensions, $\ell \cdot h$, and does not have the versatility to encode several bosonic modes \cite{Lemmer_2018}. This implies that for square lattices where \(\ell = h\), both platforms exhibit similar scaling. However, in elongated lattices, the SC circuit architecture maintains a lower circuit depth, which can lead to improved performance and reduced error accumulation.

To demonstrate the performance of our model, we consider a two-site HH model at half-filling. We perform noise-based simulations considering the current state of the art, where the DAQC approach exhibits superior performance in comparison to the digital. For various coupling regimes that can be explored with the device, the model can achieve high fidelity values, larger than 0.98 in all cases. 

Additionally, to show the relevance of our model for condensed matter systems and its applicability in simulation of materials, we explored the dynamics of a repulsively bound electron pair in a two-site system, where the coupling to phonons introduces a dissipation channel and leads to pair breaking.

Our encouraging results suggest that these types of co-design encoding are suitable to outperform classical computation capabilities in the NISQ era, making our current quantum technology useful for industrial problems related to condensed-matter physics.

\section{Methods}
\subsection{Activating digital and analog blocks}
\label{sec.A}
We now focus on mimicking each interaction term of the target Hamiltonian of Eq.~(\ref{Eq04}). The on-site repulsion $\sigma_{2j}^z\sigma_{2j-1}^z=\sigma_{L,j}^z\sigma_{R,j}^z$ and the e-p terms can be realized simultaneously by turning off all the magnetic fluxes ($A_{QQ}=A_{QR}^{(j)}=\bar{A}_{QQ}=0$), applying a Hadamard gate over each qubit, and then turning on the signal $QQ$ and $QR$ in each building block ($A_{QQ}\ne0$ and $A_{QR}\ne0$), while adjusting $\tilde{\varphi}_{QQ}^+=\tilde{\varphi}_{QQ}^-=0$. This allows the system to evolve according to
\begin{equation}
    U_{\star}(t)=e^{-it\sum_j[A_{zz}\sigma_{L,j}^x\sigma_{R,j}^x+A_{e-ph}(\sigma_{L,j}^x+\sigma_{R,j}^x)(a_j+a_j^{\dagger})]}
\end{equation}
with $A_{zz}\sim A_{QQ}$ and $A_{e-ph}\sim A_{QR}$. The operators $\sigma_{R(L),j}^x$ act on the right (left) qubit at the same site~$j$. Finally, we again apply Hadamard gates over all qubits, turning off all the signals. Accordingly, the implementation of the on-site interaction and electron-phonon coupling terms can be implemented as 
\begin{eqnarray}
 e^{-it(H_{c}+H_{e-ph}})=H^{\otimes 2n} U_{\star}(t) H^{\otimes 2n}  ,
\end{eqnarray}
where $H^{\otimes 2n}$ are Hadamard gates applied over each qubit. 

For the fermionic part, we consider a HH model with a $\ell\times h$ lattice with $\ell\cdot h=N$ and $h>\ell/2$, implying a spinless fermion lattice of $2\ell\times h$ sites. We note that in such a geometry, there are horizontal and vertical nearest-neighbor sites. For the horizontal ones, $l=j+1$. In Eq.~(\ref{Eq04}), we then have the fermionic term
\begin{eqnarray}
H_{\leftrightarrow} = \!\!\!\!\! & \sum_{j=1}^{N}(\sigma_{j,L}^x\sigma_{j,R}^z\sigma_{j+1,L}^x + \sigma_{j,R}^x\sigma_{j+1,L}^z\sigma_{j+1,R}^x)\nonumber \\
& + \sum_{j=1}^{N} (\sigma_{j,L}^y\sigma_{j,R}^z\sigma_{j+1,L}^y + \sigma_{j,R}^y\sigma_{j+1,L}^z\sigma_{j+1,R}^y) .
\label{Eq09}
\end{eqnarray}
Here, the first part can be written as
\begin{eqnarray}
    U_{xx} \sum_{j=1}^{N}(\sigma_{j,L}^x\sigma_{j,R}^y + \sigma_{j+1,L}^y\sigma_{j+1,R}^x) U_{xx}^{\dagger} ,
    \label{Eq10}
\end{eqnarray}
where the gate is $U_{xx}=e^{-i\frac{\pi}{4}\sum_j\sigma_{j,R}^x\sigma_{j+1,L}^x}$. Such nearest neighbour Ising interactions are natively available in superconducting hardwares \cite{Pedram.2024}. Moreover, these interactions can be implemented via the evolution generated by the device Hamiltonian in Eq.~(\ref{Eq06}), setting $\bar{A}_{QQ}\ne 0$, $\tilde{\phi}_{QQ}^+=0$, and $\tilde{\phi}_{QQ}^-=\pi$ for a duration of $t=\pi/(m_1 \tilde{A}_{QQ})$. Consequently, the terms from the first line of Eq.~(\ref{Eq09}) can be simulated by the gate sequence $U_{xx} \, U_{xy}(t) \, U_{xx}^{\dagger}$, where the operation $U_{xy}(t)$ can be generated from the device Hamiltonian by setting $A_{QQ}=\bar{A}_{QQ}\ne 0$ and $\tilde{\varphi}^+_{QQ}=\tilde{\varphi}^-_{QQ}=-\pi/2$. The last gate of the sequence, $U_{xx}^{\dagger}$, is implemented by setting $\tilde{\phi}^+_{QQ}=-\tilde{\phi}^-_{QQ}=\pi/2$.

In the same way, the second line in Eq.~(\ref{Eq09}) can be implemented as
\begin{eqnarray}
    U_{yy} \sum_{j=1}^{N}(\sigma_{j,L}^y\sigma_{j,R}^x + \sigma_{j+1,L}^x\sigma_{j+1,R}^y) U_{yy}^{\dagger} ,
    \label{Eq12}
\end{eqnarray}
where $U_{yy} = e^{-i\frac{\pi}{4}\sum_j\sigma_{j,R}^y\sigma_{j+1,L}^y}$, which can be implemented by the evolution generated by the device Hamiltonian of Eq.~(\ref{Eq06}) with $\bar{A}_{QQ}\ne 0$ and $\tilde{\phi}_{QQ}^+=\tilde{\phi}_{QQ}^-=0$ over a duration $t = \pi / (m_1 \tilde{A}_{QQ})$. Thus, the terms present in the second line of Eq.~(\ref{Eq09}) can be simulated by the gate sequence $U_{yy} \, U_{yx}(t) \, U_{yy}^{\dagger}$, where the operation $U_{yx}(t)$ can be implemented by setting $A_{QQ}=\bar{A}_{QQ}\ne 0$, where $\tilde{\varphi}^+_{QQ}=-\tilde{\varphi}^-_{QQ}=\pi/2$ and $\tilde{\phi}^+_{QQ}=\tilde{\phi}^-_{QQ}=-\pi/2$.

\begin{figure*}
\centering
\includegraphics[width=0.95\linewidth]{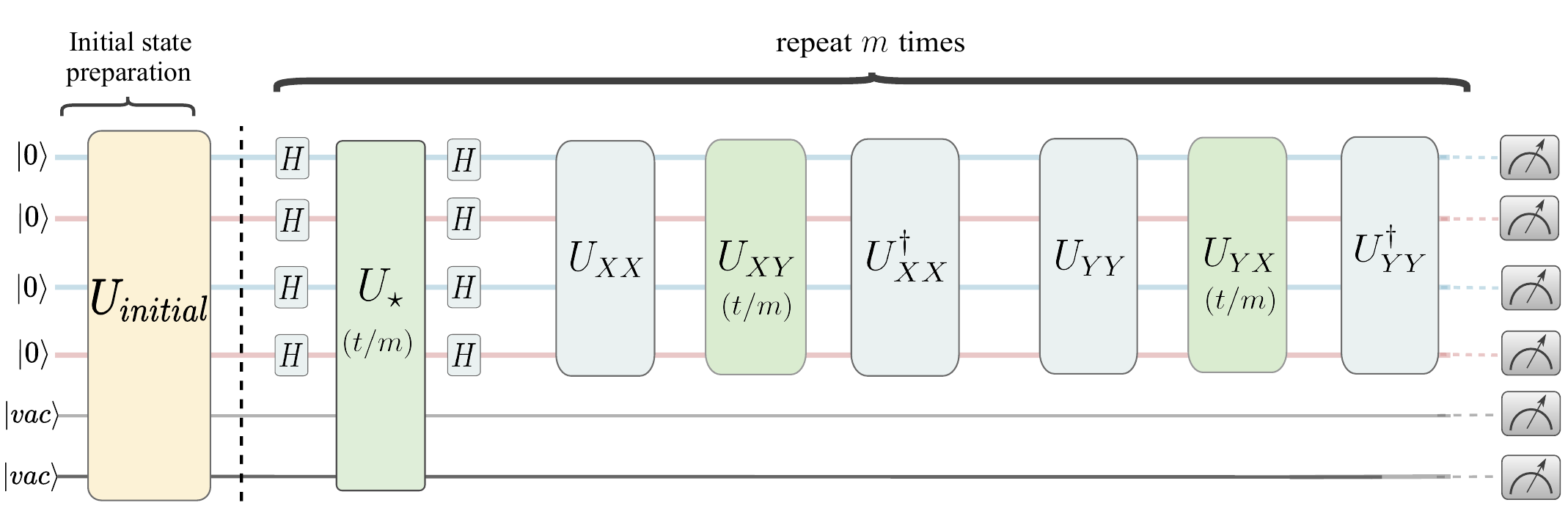}
\caption{DAQC circuit for a 2-site HH model. The gates in green are the analog blocks while the blue ones are the digital steps. The circuit depth is $9$ which does not scale as we add more sites in the 1D array.}
\label{Fig02}
\end{figure*}

For the vertical fermionic terms $l=\ell +j$, we have
\begin{eqnarray}
H_{\updownarrow} = \sum_{j=1}^{N} \sigma_{j,L(R)}^x \sigma_{j,R(L)}^z \left( \prod_{l=j+1}^{\ell+j-1} \sigma_{k,L(R)}^z \sigma_{k,R(L)}^z \right) \sigma_{\ell+j,L(R)}^x \nonumber \\
 + \sum_{j=1}^{N} \sigma_{j,L(R)}^y \sigma_{j,R(L)}^z \left( \prod_{l=j+1}^{\ell+j-1} \sigma_{l,L(R)}^z \sigma_{l,R(L)}^z \right) \sigma_{\ell+j,L(R)}^y. \nonumber \\
\label{Eq14}
\end{eqnarray}
This term is more complex and to implement it, we can follow a similar approach to Ref.~\cite{Jing.2021}. Without loss of generality, we consider odd $\ell$ and write the first term of Eq. (\ref{Eq14}) as
\begin{eqnarray}
&&\sigma_{j,L(R)}^x\sigma_{j,R(L)}^z\bigg(\prod_{l=j+1}^{\ell+j-1}\sigma_{l,L(R)}^z\sigma_{k,R(L)}^z\bigg)\sigma_{\ell+j,L(R)}^x\nonumber\\
=&&\mathcal{U}_{xy}^{j,L(R)} u_{xx}^{j,L(R)}\sigma_{s_j,L(R)}^{x}\sigma_{s_j,R(L)}^{y}u_{xx}^{j,L(R)\dagger}\mathcal{U}_{xy}^{j,L(R)\dagger}.\quad
\label{Eq15}
\end{eqnarray}
Here, $s_j=(\ell-1)/2+j$, 
\begin{equation*}
    u_{xx}^{j,L}=e^{-\frac{\pi}{4} \sigma_{s_j,R}^{x}\sigma_{s_j+1,L}^{x}},
\end{equation*}
and
\begin{eqnarray}
\mathcal{U}_{xy}^{j,L}=&\prod_{l=0}^{(\ell-3)/2}\left[e^{-1\frac{\pi}{4}(\sigma^x_{j+l,L}\sigma^y_{j+l,R}+\sigma^y_{j-l+\ell-1,R}\sigma^x_{j-l+\ell,L})}\right]\nonumber \\
&\cdot \left[e^{-1\frac{\pi}{4}(\sigma^x_{j+l,R}\sigma^y_{j+l+1,L}+\sigma^y_{j-l+\ell-1,L}\sigma^x_{j-l+\ell-1,R})}\right].\nonumber \\
\label{Eq16}
\end{eqnarray}

\subsection{Circuit depth}
To implement terms of the form given by Eq.~(\ref{Eq15}), we need to perform $2\ell$ rotations plus the interaction $\sigma^x\sigma^y$, obtaining a circuit depth of $2\ell+1$. As we have shown, all these two-body rotations can be done by adjusting the phases of the different signals (see Supplementary Table 1). In a similar way, we can write the second term of Eq.~(\ref{Eq14}) as
\begin{eqnarray}
\mathcal{U}_{yx}^{j,L} u_{yy}^{j,L}\sigma_{s_j,L}^{y}\sigma_{s_j,R}^{x}u_{yy}^{j,L\dagger}\mathcal{U}_{yx}^{j,L\dagger} ,
\label{Eq16}
\end{eqnarray}
which adds $2\ell+1$ layers to the circuit depth. We note that with $4\ell+2$ layers, we can do one vertical interaction. Nevertheless, it can be shown that all the vertical interactions can be done at the same time for the same column. This is because all the interactions commute, meaning that they can be done together as in the proposal of Ref.~\cite{Jing.2021}. Therefore, we need $2\ell(2\ell+1)$ layers to simulate all the vertical interactions. It is important to highlight that this is the only term that provides scaling to our algorithm.

The circuit depth of our algorithm is now three for all the e-p and on-site interaction terms, six for all the horizontal fermionic terms, and $2\ell(2\ell+1)$ for all the vertical ones. This means we need a total circuit depth of $2\ell(2\ell+1)+9$ to simulate a $h\times\ell$ HH lattice, where $\ell\le h$ and $\ell>1$. For the case of a chain ($\ell=1$), the circuit depth is $9$ and does not scale with the chain size. As an example, we show the circuit for our DAQC encoding for a 2-site HH model in Fig.~\ref{Fig02}. Therefore, if we want to simulate a quantum evolution of the HH model using our DAQC approach, we have, in the interaction picture,
\begin{eqnarray}
U(t)&=&e^{-it(H_{c}+H_{e-ph}+H_{\leftrightarrow}+H_{\updownarrow})}\nonumber\\
&\approx &\left[e^{-i(t/N)(H_{c}+H_{e-ph})}e^{-i(t/n)H_{\leftrightarrow}}e^{-i(t/n)H_{\updownarrow}}\right]^N . \quad
\end{eqnarray}
Here, $N$ is the number of Trotter steps and each exponential can be implemented as mentioned before. For such evolutions, our circuit depth is given by $2N\ell(2\ell+1)+9$. It is particularly interesting if we compare our findings with the digital approach, where the circuit depth always scales with both dimensions. We also note that in a digital quantum computer, the implementation of bosonic modes requires $\log_2(N)$ qubits, where $N$ is the maximum energy level considered in the bosonic dynamics. This makes our DAQC approach also efficient for in terms of hardware resources. Another platform that offers similar possibilities is trapped ions~\cite{Solano.2012}. Nevertheless, it fails in the ability to implement multiple bosonic modes and, as it uses M{\o}lmer and S{\o}rensen gates, the circuit depth also scales with the two dimensions of a lattice.

\section{Data availability}

The datasets generated and/or analyzed during the current study are available from the corresponding author upon reasonable request. Publicly available data used in this study can be accessed via \href{https://github.com/juanfrh7/daqc/tree/main}{\textbf{GitHub Repository}}.

\section{Code availability}
Due to confidential reasons, codes are available upon reasonable request.

\section{Acknowledgments} We are thankful for the financial support of Agencia Nacional de Investigaci\'on y Desarrollo (ANID): Subvenci\'on a la Instalaci\'on en la Academia SA77210018,  Fondecyt Regular 1231174, Financiamiento Basal para Centros Cient\'ificos y Tecnol\'ogicos de Excelencia AFB220001. We thank Prof. Lin Tian from University of California, Merced for fruitful discussions regarding this project.

\section{Author Contributions}

S. Kumar conceptualized the study, designed the framework, collected and analysed the results; N. N. Hegade conducted to the primary design of the framework and data analysis, and developed the quantum algorithms and simulations; A.-M. Visuri and B.A. Bhargava contributed to the theoretical framework and mathematical modeling of the physical system and its applications; J. F. R. Hernandez assisted in developing the noisy simulation model; E. Solano provided expertise in quantum computing and contributed to the conceptualization and interpretation of results; F. Albarrán-Arriagada assisted in the development of superconducting circuits model and the experimental setup; and G. Alvarado Barrios assisted in simulation procedures, data interpretation and supervised the project.

\section{Competing interests}

The authors declare no competing interests.
\newpage

\bibliography{ref} 

\end{document}